\documentstyle[11pt,aaspp4]{article}

\lefthead{Djorgovski et al.}
\righthead{Collapsed Cores in Globular Clusters}

\begin{document}

\title{Interferometric $^{12}$CO Observations of the Central Disk of NGC 4631:
An Energetic Molecular Outflow}

\author{Richard J. Rand}
\affil{Dept. of Physics and Astronomy, University of New Mexico, 800 Yale
Blvd, NE, Albuquerque, NM 87131}



\begin{abstract}

We present interferometric observations of CO $J=1-0$ emission in the
central regions of the edge-on galaxy NGC 4631, known for its extended
gaseous halo and its tidal interactions.  Previous single-dish
observations revealed that almost all of the CO emission arises from a
central ring or bar-like structure of length $\sim$ 4 kpc.  We confirm
this structure at higher resolution, and find that it is bent at the
center, reflecting the overall bend in this galaxy apparent from
optical images.  The kinematic evidence favors a rigidly rotating ring
over a bar.  The gaseous halo emission in several tracers is
concentrated above and below this molecular structure.  To the north
of an emission peak at the eastern end of the structure is an
extraplanar feature showing filamentary and shell-like properties
which we interpret as an energetic molecular outflow.  The energies
involved are difficult to estimate, but are probably of order
10$^{54}$ ergs or more.  The CO concentration in the disk below this
structure coincides with a bright HII region complex, a peak of radio
emission, and the brightest X-ray feature in the inner disk of the
galaxy seen in a ROSAT HRI map, all suggesting intense star formation.
A filament of radio continuum emission may also have a footprint in
this region of the disk.  The origin of the outflow is unclear.

\end{abstract}




%

\section{Introduction}

NGC 4631 is one of the best examples of a nearby edge-on galaxy with a
star formation driven outflow resulting in a bright and extended
gaseous halo.  This is seen in many tracers: H$\alpha$ (Rand,
Kulkarni, \& Hester\markcite{} 1992; Hoopes, Walterbos, \& Rand
\markcite{} 1999), radio continuum (e.g. Ekers \& Sancisi\markcite{}
1977; Hummel \& Dettmar\markcite{} 1990; Golla \& Hummel\markcite{}
1994), and X-rays (Wang et al.\markcite{} 1995; Vogler \& Pietsch
\markcite{} 1996).  This galaxy is undergoing interactions with two
companions, NGC 4627 and NGC 4656, and shows a complex pattern of
tidal features seen in HI (Weliachew, Sancisi, \& Gu\'elin\markcite{}
1978; Rand\markcite{} 1994), and more recently to a lesser extent in
cold dust (Neininger \& Dumke\markcite{} 1999).  It also shows two
large and highly energetic HI supershells (Rand \& van der Hulst
\markcite{} 1993), the origin of which is not clear (Rand \& Stone
\markcite{} 1996; Loeb \& Perna\markcite{} 1998).  The disturbances
to the disk resulting from the interactions may have caused the rather
high level of star formation [$L_{H\alpha} = 1.6 \times 10^{41}$ ergs
s$^{-1}$, uncorrected for internal extinction (Rand, Kulkarni \&
Hester\markcite{} 1992)].

The above-mentioned gaseous halos are particularly bright above the
central 4 kpc of the disk, where almost all of the $^{12}$CO $J=1-0$
emission is found (Golla \& Wielebinski\markcite{} 1994; hereafter
GW).  The CO observations show a rigidly rotating inner disk with a
total molecular mass of $9\times 10^8$ M$_{\sun}$.  Whether the inner
4-kpc structure might be a bar is a subject of debate (e.g. Roy, Wang,
\& Arsenault\markcite{} 1991; GW).  It is conceivable that the
interactions have driven much of the dense ISM into the central
regions of the galaxy, providing the fuel for the strong star
formation activity.

NGC 4631 provides us with an excellent opportunity to study star
formation, galaxy interactions, possible radial inflows, and the
production of gaseous halos, and thus deserves continued
investigation.  Therefore, to probe further the structure and
kinematics of the central concentration of dense gas, we have observed
it in the $^{12}$CO $J=1-0$ line with the Berkeley-Illinois-Maryland
Array (BIMA).

\section{Observations}

The data were taken in September 1995 in the C configuration of the
BIMA array, with baselines ranging from 12 to 52 m.  A single track
was obtained, with an approximate on-source integration time of five
hours.  The primary beam of the antennae is 100'' at the 2.6 mm
wavelength of the $^{12}$CO $J=1-0$ transition.  The quasar 3C273 was
used for flux calibration.  While observing NGC 4631, pointing was
switched every 50 seconds between two positions with offsets of 25''
east and 15'' west of a position near the radio nucleus [R.A. 12$^{\rm
h}$ 42$^{\rm m}$07.7$^{\rm s}$, Dec. 32$^{\circ}$ 32' 29'' (2000.0);
Duric, Seaquist, \& Crane\markcite{} 1982), ensuring that the all of
the entire bright emission from the central regions mapped by GW was
contained within the primary beams.  Calibration, mapping, and
cleaning were carried out with the MIRIAD package (Sault et
al.\markcite{} 1995), using the 'mosaic' option in the program
'invert' to create a mosaic dirty map of the two fields.  The emission
was bright enough to allow self-calibration to be performed.  For each
pointing a clean map was initially produced and used as a model in the
MIRIAD program 'selfcal'.  The self-calibrated $uv$-data were then
used to create a new clean map, which became the new model for the
next iteration.  Convergence was achieved after four iterations.  The
clean beam has dimensions 9.8''x6.7'' (P.A. --82$^{\circ}$).  The
noise in the final channel maps is 130 mJy (beam)$^{-1}$ or 0.2 K.  A
total intensity map was formed from emission in the range
$V_{lsr}=440-820$ km s$^{-1}$, using data above 2$\sigma$ in each
channel.  The systemic velocity of NGC 4631 is $V_{lsr,sys} = 618$ km
s$^{-1}$ (Rand\markcite{} 1994), and we adopt the commonly used
distance of 7.5 Mpc (Hummel, Sancisi, \& Ekers\markcite{} 1984).

\section{Results}

\subsection{General Structure}

Figure 1 shows the map of total CO intensity, including the FWHM of
the primary beam for the two pointings, while Figure 2 shows the same
map overlaid on H$\alpha$ and red-continuum CCD images from Rand,
Kulkarni, \& Hester (\markcite{}1992).  Some emission at the western
end of the CO distribution is beyond the FWHM of the western primary
beam and may not be real or accurately mapped.  Other isolated
features above the plane which reach only the first contour level
(2$\sigma$) are not significant enough to be considered real.
Position-velocity diagrams parallel to the major axis of the galaxy
are shown in Figure 3.  The major axis position angle of 86${^\circ}$ is
the large-scale value derived from HI data (Rand\markcite{} 1994).

\placefigure{fig1}
\placefigure{fig2}
\placefigure{fig3}

The basic CO structure and kinematics of the central region have been
analyzed in detail by GW; hence, we only briefly discuss them here,
and focus on the small-scale features revealed by the high resolution
(\S 3.2).  The emission shows a continuous structure with maxima at
either end.  The central minimum is at the position of the HI
kinematic center (Rand\markcite{} 1994).  The brightening towards the
far ends of these features is consistent with previous single-dish
observations (GW; Sofue et al.\markcite{} 1990; Sofue, Handa, \&
Nakai\markcite{} 1989), and suggests a ring or bar-like structure of
length $\sim$ 4 kpc.  The spectra of GW show that the emission falls
off sharply beyond the radius of this structure.  The ``bend'' noted
by GW is even more noticeable here: the eastern half is elongated
along a position angle $\approx 90{^\circ}$ while the position angle
of the western half is $75-80{^\circ}$.  In this way, the central
region reflects the well-known overall bend of the galaxy.  This bend
causes an exaggeration of the E-W asymmetry of the emission in the
position-velocity ($l-v$) diagrams: with the northern (southern)
slices showing brighter emission from the eastern (western) end.  The
midplane $l-v$ diagram indicates rigid rotation with a slope of 1.9 km
s$^{-1}$ arcsec$^{-1}$, in agreement with the slope measured from the
CO 2--1 map of GW.  Velocities reach about 130 km s$^{-1}$ from the
systemic velocity.

The bright concentration at the eastern end of the emission shows a
two-peaked velocity profile in a few cuts in Figure 3.  The
concentration at the western end does not show a clear bifurcation,
but the velocity width is relatively large, similar to the eastern
concentration.  These kinematics could be due to localized events (see
\S 3.2), or they may indicate non-circular motions.  In particular, at
the ends of a bar, gas streaming nearly perpendicular to the bar -- on
nearly circular orbits where the spiral pattern begins -- may run into
gas streaming along the bar (e.g. Regan, Vogel, \& Teuben\markcite{}
1997), creating a shock (see also Roberts, Huntley, \& van Albada
\markcite{} 1979) or at least orbit crowding.  The CO velocity field
at the transition between the bar and the spiral arms in M83 has been
interpreted in this way (Kenney \& Lord\markcite{} 1991).  This flow
convergence may lead to bright regions of star formation, as observed
at the ends of many bars, although more often in SBb galaxies
(Phillips\markcite{} 1993).  A high level of star formation at this
CO peak is indicated by bright H$\alpha$, X-ray and radio continuum
emission, as discussed in \S 3.2.

Against the bar hypothesis is the fact that nothing else in the $l-v$
diagram strongly suggests a bar, such as the expected
``parallelogram'' appearance of the emission due to $x_1$ orbits
parallel to the bar potential (Binney et al.\markcite{} 1991).  This
conclusion was also reached by GW.  Only for a bar viewed side-on do
the $x_1$ orbits occupy a narrow locus in $l-v$ diagrams (see Figure 8
of Garc\'ia-Burillo \& Gu\'elin\markcite{} 1995).  In support of such
a perspective is that NGC 4631 has one of the most slowly rising CO
rotation curves of the 14 galaxies studied by Sofue et
al. (\markcite{}1997), as might be expected for a side-on view of a
bar.  However, the low slope of the rotation curve could simply be due
to the almost complete lack of a visible bulge.  Furthermore, there is
no indication of gas on $x_2$ orbits perpendicular to the bar, deep in
the potential (cf. Figure 1 of Binney et al.\markcite{} 1991).  These
are particularly noticeable in side-on views.  A lack of molecular gas
in the inner parts of the bar could be responsible.  Alternatively,
the bar may not have an ILR (e.g. Athanassoula\markcite{} 1992).  For
the modeled HI rotation curve of Rand (\markcite{}1994), this would
imply a bar pattern speed of $\gtrsim 20$ km s$^{-1}$ kpc$^{-1}$.
Hence, a bar cannot be ruled out but there is little evidence to
support anything more complex than a ring structure.  Nevertheless,
even a ring may represent a resonant response to a bar potential, but
again there is little supporting evidence for a bar, particularly in
the stellar emission (see below).  There may still be lower-level
non-circular motions, however, given that slightly non-axisymmetric
bulges are common in spirals (Zaritsky \& Lo\markcite{} 1986), and
given the many other disturbances to this galaxy.

HI data in this central region show a rise in velocity very similar to
the CO data, but no evidence of a central minimum.  It is more
difficult to discern a bar in the HI data because of confusion with
the larger, differentially-rotating disk.  There is evidence of
broadened velocity profiles in the central 4'.  The model HI cube of
Rand (\markcite{}1994) featured a velocity dispersion dictated by the
rate of falloff of emission at the terminal velocity in the
differentially rotating component.  This choice produced generally
broader profiles than observed in the central 4', most likely
indicating non-circular motions, but whether they are due to an inner
bar or to spiral structure cannot be determined.

The disturbed optical appearance in the central regions, the lack of a
discernable bulge, and the lack of near-IR imaging of this galaxy make
it difficult to appeal to the stellar distribution to aid in
discerning a bar.  Such a structure might be inferred from a boxy or
peanut-shaped bulge, as in, for example, NGC 4013, where the CO
distribution clearly shows evidence for bar-driven orbits
(Garc\'ia-Burillo, Combes, \& Neri\markcite{} 1999).  In NGC 4631,
the central regions in the red image (Figure 2b) show a very irregular
morphology, due to some combination of irregular stellar and dust
distributions.

The central regions of NGC 4631 have been recently mapped in
submillimeter continuum emission by three different groups (Braine et
al.\markcite{} 1995; Alton, Davies, \& Bianchi\markcite{} 1999;
Neininger \& Dumke\markcite{} 1995).  The CO map in general resembles
these maps rather well in that they all show a central structure of
about 4 kpc extent with emission enhancements near the ends.  The
prominent extraplanar feature seen in the 1.3mm continuum map of
Braine et al.  (\markcite{}1995), extending about 1' southward of the
eastern complex, is not seen in CO.  It is not likely to have been
resolved out, since essentially all of its structure is on scales to
which the CO map is sensitive.  Using the conversion of 1.3mm surface
brightness to gas mass given by Braine et al., the typical gas surface
density in this filament should be of order 200 M$_{\sun}$ pc$^{-2}$.
If all of this gas were molecular and detectable in CO, we should
expect emission in our map at a typical level of 30 Jy (beam)$^{-1}$
km s$^{-1}$, or 6$\sigma$.  This is clearly not seen.  Possibilities
are that the dust is associated with HI gas, the gas to dust ratio is
much lower in NGC 4631, any molecular gas in the spur is too cold to
emit substantially in the CO lines, or the emission is spurious.  For
the main central 2' feature at least, the agreement of the derived
mass with the total HI+H$_2$ mass is better than a factor of 2 (Braine
et al.), suggesting that the essential conversion from 1.3mm emission
to total gas mass is reasonably accurate.  It is difficult to extend
this conclusion to the high-$z$ gas, however.  More relevant is that
the feature is not seen in the 1.2mm map of Neininger \& Dumke
(\markcite{}1999), which has very similar resolution and sensitivity.
It may well be spurious.

On the other hand, the beginnings of the loop south of the nucleus in
Braine et al.'s map coincide with a CO spur extending south from the
western half of the emission; these in turn may be associated with an
extraplanar H$\alpha$ feature (Figure 2).  The 850 $\mu$m continuum
map of Alton et al. (\markcite{}1999) also shows extraplanar emission
just south of this CO spur.

The 1.4 GHz map of Golla (\markcite{}1999) at 1.45'' resolution shows
much agreement with the CO distribution on scales of a few arcseconds,
including bright concentrations coincident with the two CO peaks at
each end of the main structure, and a relative deficit of emission
between them.  Earlier radio maps also show a concentration of
synchrotron emission in the central 4 kpc of the disk (e.g. Ekers \&
Sancisi\markcite{} 1977, Duric et al.\markcite{} 1982; Golla \&
Hummel\markcite{} 1994).

Most of the diffuse extraplanar H$\alpha$ emission (Figure 2; see also
Hoopes, Walterbos, \& Rand\markcite{} 1999) is found above and below
the 4-kpc extent of the CO emission.  The quasi-vertical double-worm
structure reported by RKH does not rise from the center of the CO
structure but from a position about 700 pc to the east.  Diffuse, soft
X-ray halo emission generally surrounds the inner disk and is
brightest in the approximate radial range of the CO emission, as seen
in the overlay of the ROSAT PSPC map from Vogler \& Pietsch
(\markcite{}1996), kindly provided by A. Vogler, and the CO map in
Figure 4a (see also Wang et al.\markcite{} 1995).  Finally, bright
polarized emission at 4.88 and 8.46 GHz is found in the halo above
this region of the disk, where the inferred intrinsic magnetic field
runs largely perpendicular to the major axis (Golla \& Hummel
\markcite{} 1994).  All of these associations suggest active star
formation and outflow of gas into the halo.  All three halo tracers
are brighter on the north side of the CO emission than the south side.

\placefigure{fig4}

Assuming a conversion between CO brightness and H$_2$ column density
of $X = 2.3 \times 10^{20}\ {\rm mol\ cm^{-2}\ (K\ km\ s^{-1})^{-1}}$,
the total molecular mass detected is $1.0\times 10^9$ M$_{\sun}$.
Within the uncertainties, this value agrees with the estimate of GW for
the central 140'' (using the same value of $X$), suggesting that
little, if any, emission has been resolved out by our observations.

\subsection{An Unusual Extraplanar Feature}

The most important new result from this observation is the filamentary
and shell-like emission found north of the eastern end of the main CO
feature.  A weaker feature is also seen to the south.  The extraplanar
emission is detected up to 750 pc above the midplane of the CO
distribution.  Although the galaxy is not quite edge-on, it is
unlikely that the feature is in the plane of the galaxy.  At
$i=86{^\circ}$, it would have to extend about 10 kpc along the
midplane to explain the apparent extent parallel to the minor axis.
Also, kinematically, in a highly inclined, differentially rotating
disk, one would expect the feature to show a smooth gradient towards
lower observed velocities with increasing distance from the major axis
if it were in the disk.  Although its kinematics are complex (see
below), it does not show such a signature.  The mass of the feature is
estimated at roughly 10$^8$ M$_{\sun}$, using the above value of $X$.
It splits into a quasi-linear feature on the west side, rising to the
north-west, and a more shell-like feature on the east side, suggesting
that it may have been produced by two events.

Figure 5 shows position-velocity diagrams parallel to the R.A. axis in
the region of this feature, covering its full vertical extent.  The
disk emission below the feature is the brightest in the map and
kinematically splits into two components, as discussed above, with the
lower velocity component deviating somewhat from the general trend of
velocity with position.  Just to the east of the bright disk feature,
most of the emission centers around $V_{lsr} \approx 750$ km s$^{-1}$,
while fainter emission is seen at $V_{lsr} \approx 680$ km s$^{-1}$
(more apparent in the $-3''$ panel).

\placefigure{fig5}

Velocities in the western, filamentary component of the extraplanar
feature (best seen in the 12, 15 and 18'' panels at R.A. offsets
around --12'') center around $V_{lsr}=680$ km s$^{-1}$, a typical
velocity for its distance along the major axis, and no clear trend of
velocity with position can be identified as a clue to its nature.
However, in the eastern extraplanar feature (around R.A. offset 12''),
the two disk components at $V_{lsr} \approx 750$ km s$^{-1}$ and
$V_{lsr} \approx 680$ km s$^{-1}$ are seen up to the 15'' panel (also
to the --6'' panel).  This velocity splitting may indicate a structure
expanding at $v\approx 70$ km s$^{-1}$.  Given that the eastern
feature accounts for slightly more than half the emission, the
equivalent kinetic energy of a spherically expanding shell would be
enormous: $2-3 \times 10^{54}$ erg, comparable to the two HI
supershells.  On the other hand, there are other possible sources of
peculiar kinematics at this radius, as discussed above, and the line
splitting may have an orbital origin, although the evidence is inconclusive.

Another indication of the energy involved in creating such an outflow
comes from an estimate of the potential energy of the gas mass at its
height above the plane.  This has been estimated for several dust
clouds above the disk of NGC 891 by Howk \& Savage (\markcite{}1997).
Unfortunately, the estimate depends on the rather poorly known total
midplane mass density and scale height.  The potential energy is given
by:
\begin{equation}
\Omega = 10^{52}\, {\rm ergs}\,({M\over 10^5 M_{\sun}})\,({z_0\over
700{\rm pc}})\,({\rho_0\over 0.185 M_{\sun} {\rm pc}^{-3}})\,{\rm
ln[cosh}(z/z_0)]
\end{equation}
where $z_0$ is the total mass scale height, $\rho_0$ is the total
midplane mass density and $M$ is the mass of the high-$z$ feature.
The $B-$band scale-height has been determined by Hummel \& Dettmar
(\markcite{}1990) to be about 1 kpc.  This may be a poor estimate of
the true mass scale height, which itself could vary significantly
around this disturbed galaxy.  However, use of this value should still
give a rough estimate of the energies involved.  The total mass
density is unknown but presumably less than in NGC 891 or the Milky
Way given the low rotation speed of 140 km s$^{-1}$ (Rand\markcite{}
1994).  Using the above molecular gas mass and an average height above
the plane of 500 pc, the potential energy is then
\begin{equation}
\Omega = 2\times 10^{54}\, {\rm ergs}\, ({\rho_0\over 0.185 M_{\sun}
{\rm pc}^{-3}})
\end{equation}
comparable to the kinetic energy estimate.  Of course, this analysis
does not account for energy radiated away or converted to thermal or
turbulent motions.

The bright disk emission below this feature abuts the western edge of
the very bright complex of HII regions CM65, CM66, and CM67 (Crillon
\& Monnet\markcite{} 1969; Roy et al.\markcite{} 1991), suggesting a
region of intense star formation.  In fact, the eastern, shell-like
extraplanar feature almost appears to surround the peak of H$\alpha$
(and associated red continuum) emission (Figure 2).  The H$\alpha$
luminosity from this complex, measured from the image of Rand,
Kulkarni, \& Hester (\markcite{}1992), is $6\times 10^{39}$ erg
s$^{-1}$ (uncorrected for extinction), about equal to that of 30
Doradus in the LMC (Kennicutt \& Hodge\markcite{} 1986).  Roy et
al. (\markcite{}1991) noted a complex velocity structure for these
HII regions, with components at $V_{lsr}$ = 641, 687, and 753 km
s$^{-1}$.  Thus, both CO and H$\alpha$ emission show a large range of
velocities, and there is some correlation of these components between
the two tracers.  Of course, it must be recalled that the disk
emission may have a significant extent along the line of sight (for
instance if it is part of a ring viewed along the tangent point), and
this may explain, at least in part, why it is so bright in several ISM
tracers.  There are no well-defined {\it extraplanar} H$\alpha$
features which are associated with the CO emission.  In particular,
the double-worm structure lies between this CO structure and the
center of the molecular ring.  It may originate from a star-forming
complex elsewhere on the ring.

X-rays from this part of the disk provide further evidence for a high
rate of star formation.  Figure 4b shows the CO emission with contours
of X-ray emission from the ``adaptive-filtered'' ROSAT HRI (0.1--2.4
keV) image of Vogler \& Pietsch (\markcite{}1996).
The filtering technique smooths the original HRI image with a kernel
whose size depends on the image intensity, thus enhancing low-level
diffuse emission while retaining high resolution for bright, compact
features.  No X-ray feature can be specifically associated with the
molecular outflow, but the bright concentration of disk CO emission
beneath it is coincident within the positional uncertainties with the
brightest peak of X-ray emission in the inner disk.  This source has a
luminosity, corrected for Galactic absorption, of $8 \pm 3 \times
10^{37}$ erg s$^{-1}$ (0.1--2.4 keV), and is too weak to obtain a
spectrum.  This luminosity is a lower limit due to unaccounted for
absorption in NGC 4631, and is comparable to those of the X-ray sources
in M101 postulated to be hypernova remnants by Wang (\markcite{}1999).

Coincident with these disk CO and X-ray sources is a bright source of
radio emission in Golla's (\markcite{}1999) 1.4 GHz map at 1.45''.
This source resolves into five features in a 4.86 GHz map at
sub-arcsecond resolution, four of which have thermal spectra.  Golla
\& Hummel (\markcite{}1994) trace one of the radio continuum spurs
back down to the CM67 region in the disk.  None of the extraplanar
emission from cold dust found by Neininger \& Dumke (\markcite{}1999)
appears to be associated with the CO feature.  Examination of the HI
data cube of NGC 4631 at 12''x22'' (Rand\markcite{} 1994) shows no
obvious associated kinematic or morphological peculiarities.

Hence, there is much evidence for a very intense region of star
formation at in this disk location, and an extraplanar radio continuum
filament also seems to be associated with the CO feature.

\subsection{Possible Origins of the Extraplanar Feature}

It is difficult to conclude much about the origin of this feature at
this point.  If the kinetic energy is of order 10$^{54}$ ergs, then it
may be difficult to explain it as a result of stellar winds and
supernovae, as was the case for the two large HI supershells (Rand \&
van der Hulst\markcite{} 1993).  One would require of order 10,000
supernova progenitors.  Alternatively, it may be due to the impact of
a rather massive high-velocity cloud, an explanation considered by
Rand \& Stone (\markcite{}1996) for the HI supershells (there, an
impactor of about 10$^7$ M$_{\sun}$ was required).  The impact may
have triggered the star formation in the disk at this location.  In
this case, the X-ray emission would be due to the resulting
supernovae, as the collision itself would not be able to heat gas to
X-ray emitting temperatures.  An examination of the HI data of Rand
(\markcite{}1994) does not reveal any obvious signs of an impact at
this location, such as a trail of gas.  Finally, there is the newly
recognized possibility that such shells can be produced by a single
energetic explosion -- a hypernova, which may be the cause of
gamma-ray bursts (e.g. Loeb \& Perna\markcite{} 1998; Efremov,
Elmegreen, \& Hodge\markcite{} 1998; Wang\markcite{} 1999).  At best,
it is difficult to distinguish between putative hypernovae and
conventional supernova-driven supershells, and in this case especially
so, given the difficulty of disentangling emission along the line of
sight, correcting X-ray fluxes for absorption in the intervening gas,
and determining the age of the high-$z$ structure.  However, it is
worth noting that this is now the third problematically energetic
event found in this galaxy.  This may provide constraints on the
hypernova hypothesis.  Whether these events occur because of
hypernovae, or multiple supernovae and stellar winds from massive star
formation concentrated into unusually large associations, or a few
relatively massive high-velocity clouds penetrating the disk, can only
be answered by future investigation.  The CO emission from the feature
reported here is bright enough to allow a higher resolution map to be
made.  Such a map may provide a clearer delineation of its morphology
and kinematics, allowing a more careful comparison with models of its
origin.  Also Chandra, with its higher resolution compared to ROSAT,
may constrain the nature of the X-ray emission by showing whether it
remains diffuse on smaller scales or breaks up into multiple
components, and also by allowing a measurement of the gas temperature.

\section{Conclusions}

In this brief paper we have presented a high-resolution map of the
centrally concentrated CO emission in NGC 4631.  The map confirms the
ring or bar-like distribution of molecular gas inferred from
single-dish data, but there is little direct evidence for a bar.  Halo
emission in several gaseous tracers is concentrated above, and to a
lesser extent, below, this molecular structure, suggesting that the
central concentration of molecular gas has provided the fuel for
intense star formation which has resulted in a substantial outflow of
gas.

Of great interest is an extraplanar CO feature found above the eastern
end of the molecular structure.  It suggests an outflow of about
$10^8$ M$_{\sun}$ of molecular gas, although there is only weak
evidence for an associated kinematic signature.  Nevertheless, the
energies involved in driving the outflow may be well over 10$^{54}$
ergs.  Bright HII regions and X-ray emission suggest that the
underlying disk location is very active in star formation.  An
extraplanar radio continuum filament also appears to have a footprint
in this region.  Like the two highly energetic HI supershells in this
galaxy, this outflow may be driven by multiple supernovae or is
perhaps the result of an impact of a high-velocity cloud.  Its
characteristics also put it in the class of events postulated to be
due to hypernova explosions.

\footnotetext{The National Radio Astronomy Observatory is operated by
the Associated Universities, Inc. under cooperative agreement with the
National Science Foundation.}

This research has made use of the NASA/IPAC Extragalactic Database
(NED) which is operated by the Jet Propulsion Laboratory, California
Institute of Technology, under contract with the National Aeronautics
and Space Administration.
 
We are grateful to W. Pietsch and A. Vogler for providing the ROSAT
images, and to P. Teuben for comments on the manuscript.

\begin{figure}
\caption{Map of total CO intensity at $9.8 \times 6.7\arcsec$
resolution.  The dashed circles indicate the HPBW of the primary beam
for the two pointings.  The clean beam is shown at lower right.  A bar
corresponding to a length of 1 kpc is also shown.  Contour levels are
10 to 180 Jy (beam)$^{-1}$ km s$^{-1}$ in steps of 15 Jy (beam)$^{-1}$
km s$^{-1}$.  The 1$\sigma$ noise in this map is about 5 Jy
(beam)$^{-1}$.
\label{fig1}}
\end{figure}

\begin{figure}
\caption{Contours of CO emission overlaid on {\it a)} an H$\alpha$ image
of NGC 4631 showing the central $\sim$ 13 kpc of the disk, and {\it
b)} a red image showing the central $\sim$ 7 kpc of the disk (both
from Rand, Kulkarni, \& Hester 1992).  Both optical images are displayed
on a logarithmic scale.  Contour levels are as in Figure
1.  Spillover due to a bright star is present in both optical images.
\label{fig2}}
\end{figure}

\begin{figure}
\caption{Major axis position-velocity diagrams along a position angle of
86$^{\circ}$.  East is to the left.  Slices are spaced by 3''
perpendicular to the major axis.  The central slice has its origin at
R.A. 12$^{\rm h}$ 42$^{\rm m}$07.7$^{\rm s}$, Dec. 32$^{\circ}$ 32'
29''.  Contour levels are --0.3 to 2.7 Jy (beam)$^{-1}$ in steps of
0.3 Jy (beam)$^{-1}$ (about 2.3$\sigma$).
\label{fig3}}
\end{figure}

\begin{figure}
\caption{Contours of CO emission overlaid on the central part of the (a)
ROSAT PSPC image and (b) ROSAT HRI image from Vogler \& Pietsch (1996).
Contour levels are as in Figure 1.
\label{fig4}}
\end{figure}

\begin{figure}
\caption{Position-velocity diagrams along a position angle of
90$^{\circ}$ for the eastern end of the main emission.  Slices are
spaced by 3'' perpendicular to the major axis.  The central slice has
its origin at R.A. 12$^{\rm h}$ 42$^{\rm m}$10.47$^{\rm s}$,
Dec. 32$^{\circ}$ 32' 35''. Contour levels are as in Figure 3.
\label{fig5}}
\end{figure}

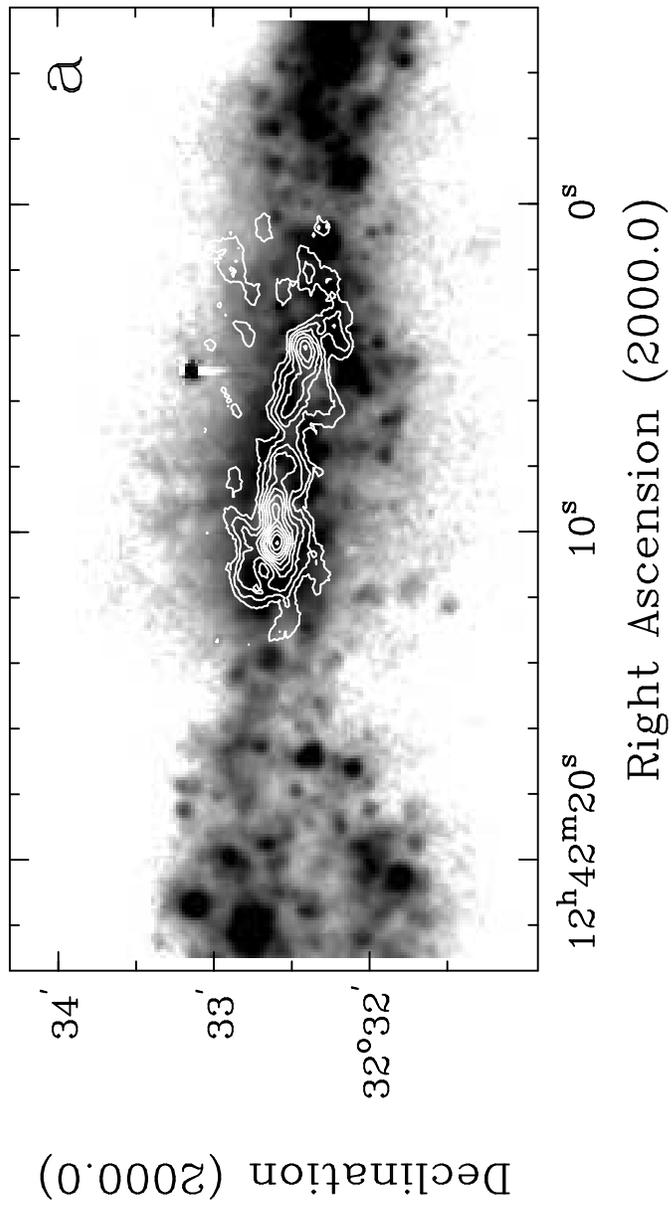
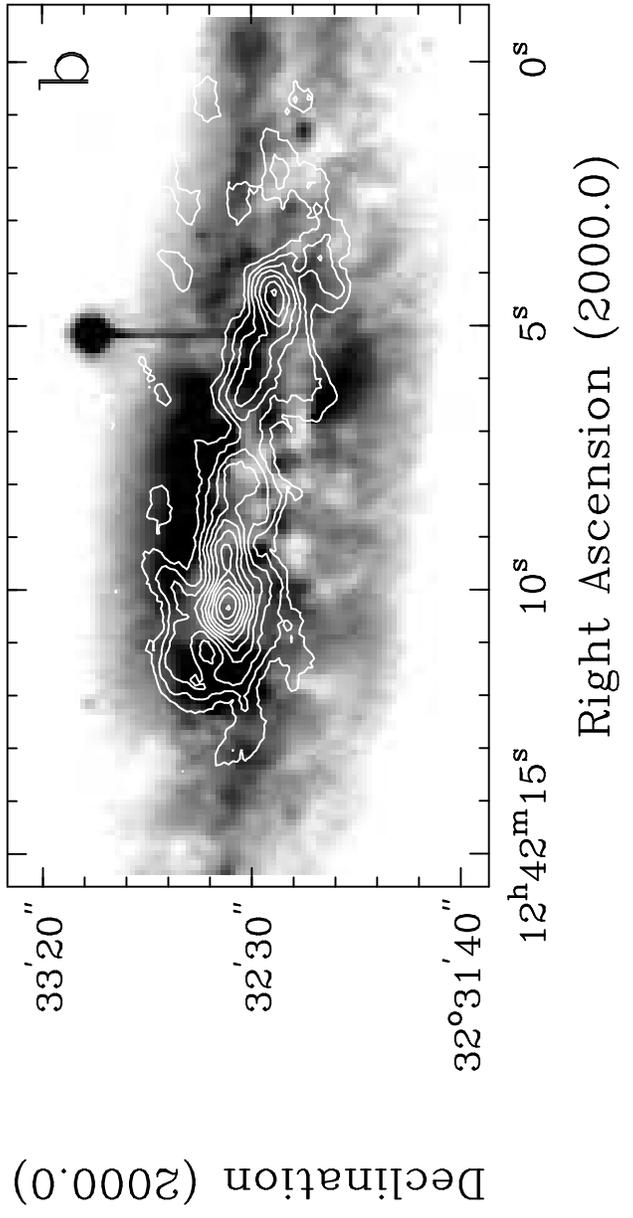
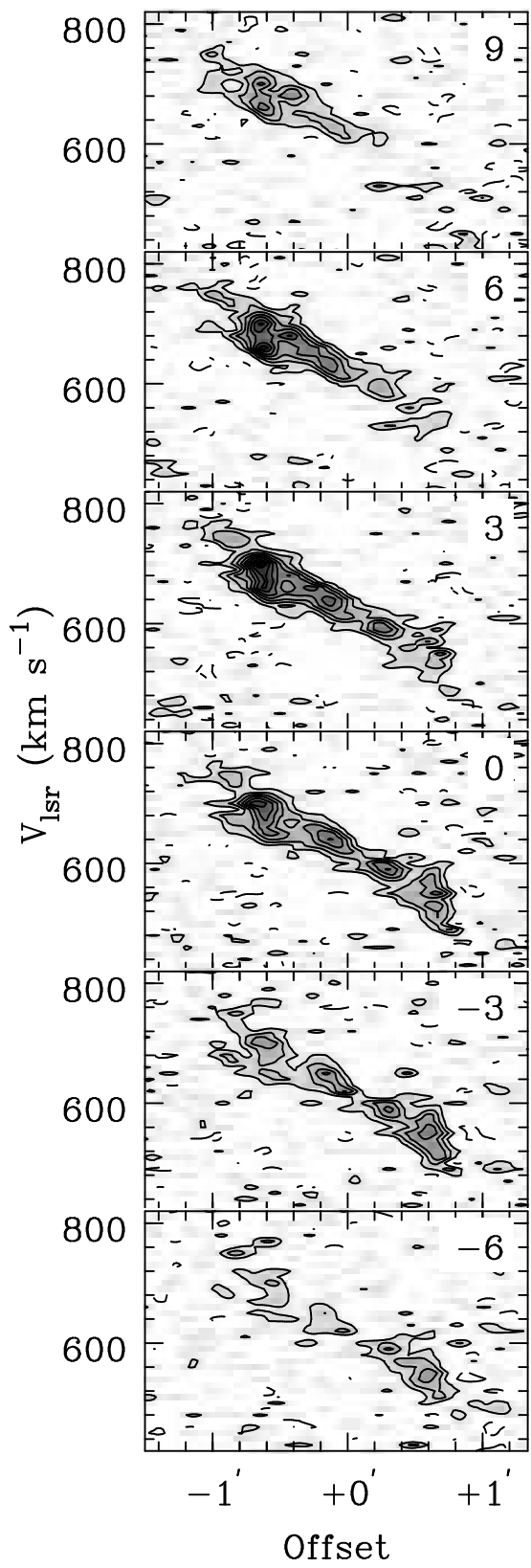
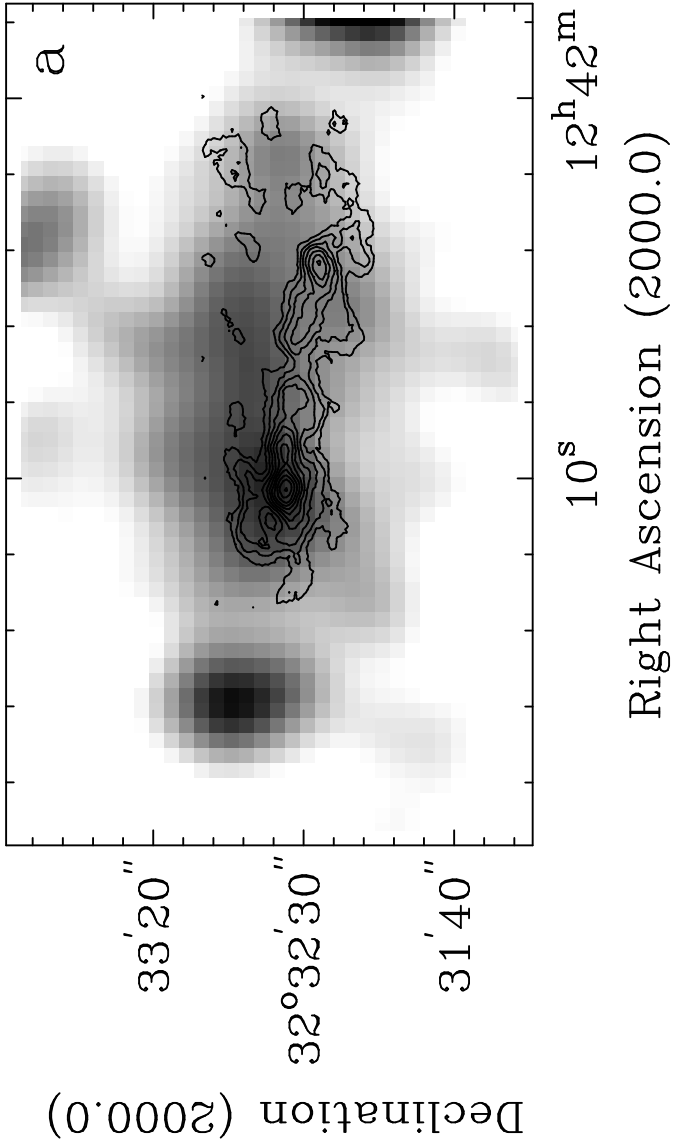
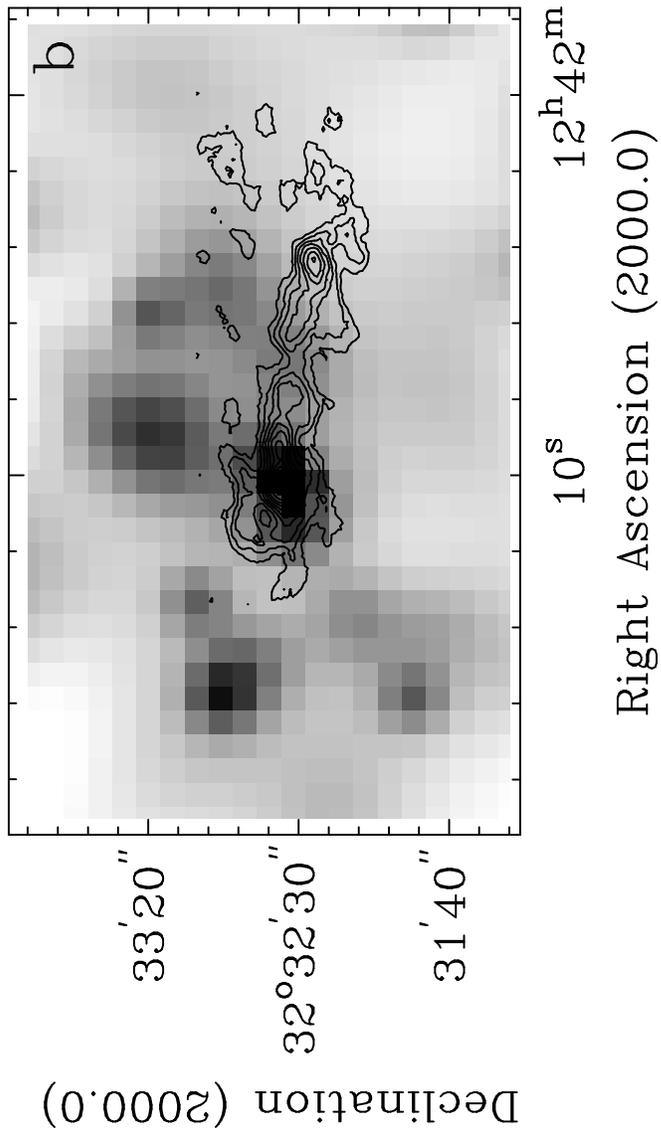
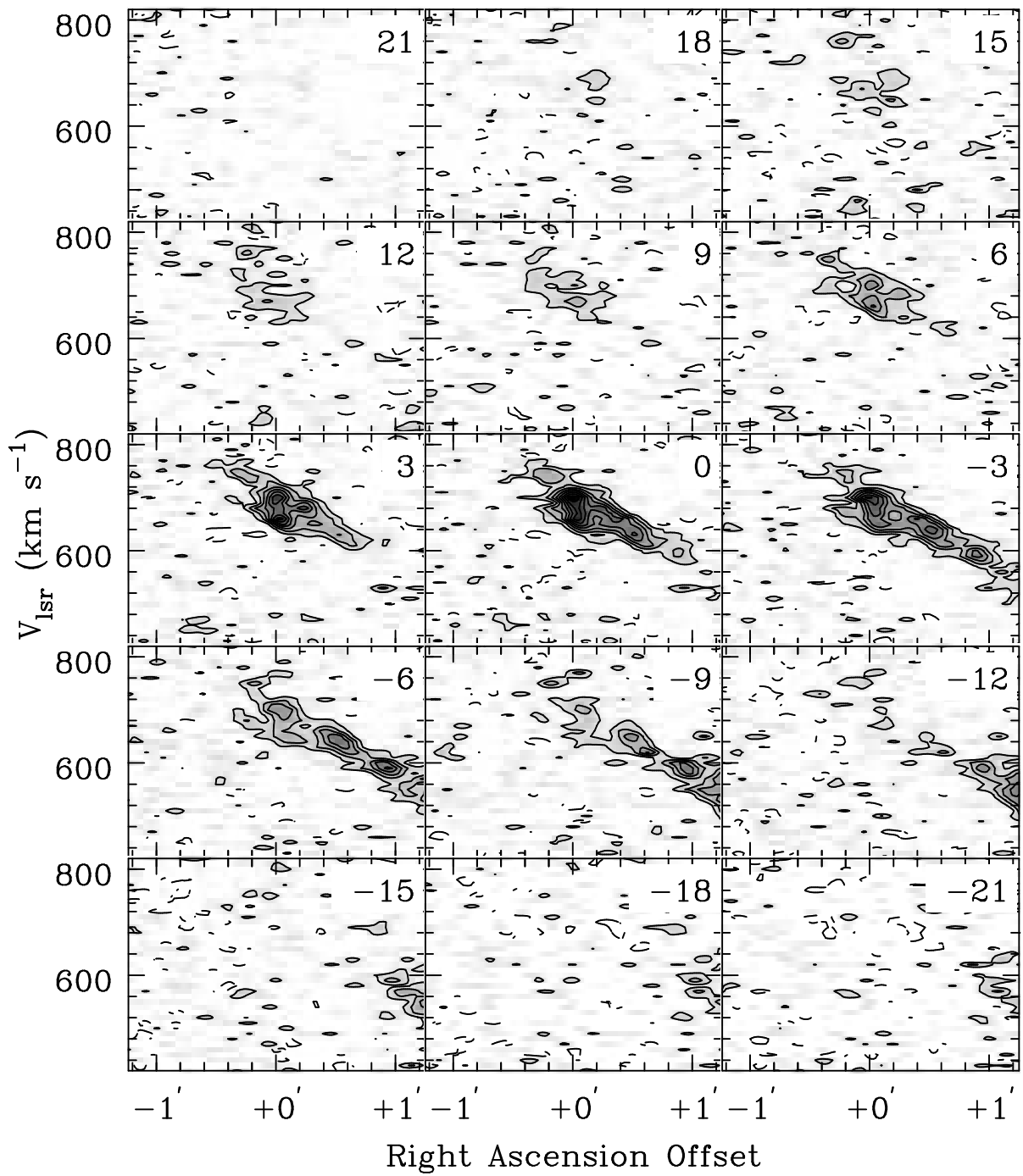
\end{document}